\documentclass[sigconf]{acmart}
\usepackage{listings}

\AtBeginDocument{%
  }

\setcopyright{acmlicensed}
\copyrightyear{2024}
\acmYear{2024}
\acmDOI{XXXXXXX.XXXXXXX}

\acmConference[Workshop on Generative AI for Recommender Systems and Personalization, KDD 2024]{August 25--26, 2024}{Barcelona, Spain}

\begin{document}

\title{Meta Knowledge for Retrieval Augmented Large Language Models}


\author{Laurent Mombaerts}
\affiliation{%
  \institution{Amazon Web Services}
   \city{Luxembourg}
   \country{Luxembourg}
 }
 \email{lmomb@amazon.com}

 \author{Terry Ding}
 \affiliation{%
   \institution{Amazon Web Services}
   \city{Arlington, VA}
   \country{USA}}
 \email{tianyd@amazon.com}

 \author{Florian Felice}
 \affiliation{%
   \institution{Amazon Web Services}
   \city{Luxembourg}
   \country{Luxembourg}
 }
 \email{flofelic@amazon.com}

 \author{Jonathan Taws}
 \affiliation{%
  \institution{Amazon Web Services}
  \city{Luxembourg}
  \country{Luxembourg}}
 \email{tawsj@amazon.com}

 \author{Adi Banerjee}
 \affiliation{%
  \institution{Amazon Web Services}
  \city{Boston, MA}
  \country{United States}}
 \email{adibaner@amazon.com}

 \author{Tarik Borogovac}
 \affiliation{%
  \institution{Amazon Web Services}
  \city{Boston, MA}
  \country{United States}}
 \email{tarikbo@amazon.com}

\renewcommand{\shortauthors}{Mombaerts, et al.}

\begin{abstract}
Retrieval Augmented Generation (RAG) is a technique used to augment Large Language Models (LLMs) with contextually relevant, time-critical, or domain-specific information without altering the underlying model parameters. However, constructing RAG systems that can effectively synthesize information from large and diverse set of documents remains a significant challenge. We introduce a novel data-centric RAG workflow for LLMs, transforming the traditional \textit{retrieve-then-read} system into a more advanced \textit{prepare-then-rewrite-then-retrieve-then-read} framework, to achieve higher domain expert-level understanding of the knowledge base. Our methodology relies on generating metadata and synthetic Questions and Answers (QA) for each document, as well as introducing the new concept of Meta Knowledge Summary (MK Summary) for metadata-based clusters of documents. The proposed innovations enable personalized user-query augmentation and in-depth information retrieval across the knowledge base. Our research makes two significant contributions: using LLMs as evaluators and employing new comparative performance metrics, we demonstrate that (1) using augmented queries with synthetic question matching significantly outperforms traditional RAG pipelines that rely on document chunking ($p < 0.01$), and (2) meta knowledge-augmented queries additionally significantly improve retrieval precision and recall, as well as the final answer’s breadth, depth, relevancy, and specificity. Our methodology is cost-effective, costing less than \$20 per 2000 research papers using Claude 3 Haiku, and can be adapted with any fine-tuning of either the language or embedding models to further enhance the performance of end-to-end RAG pipelines.
\end{abstract}



\begin{teaserfigure}
  \includegraphics[width=\textwidth]{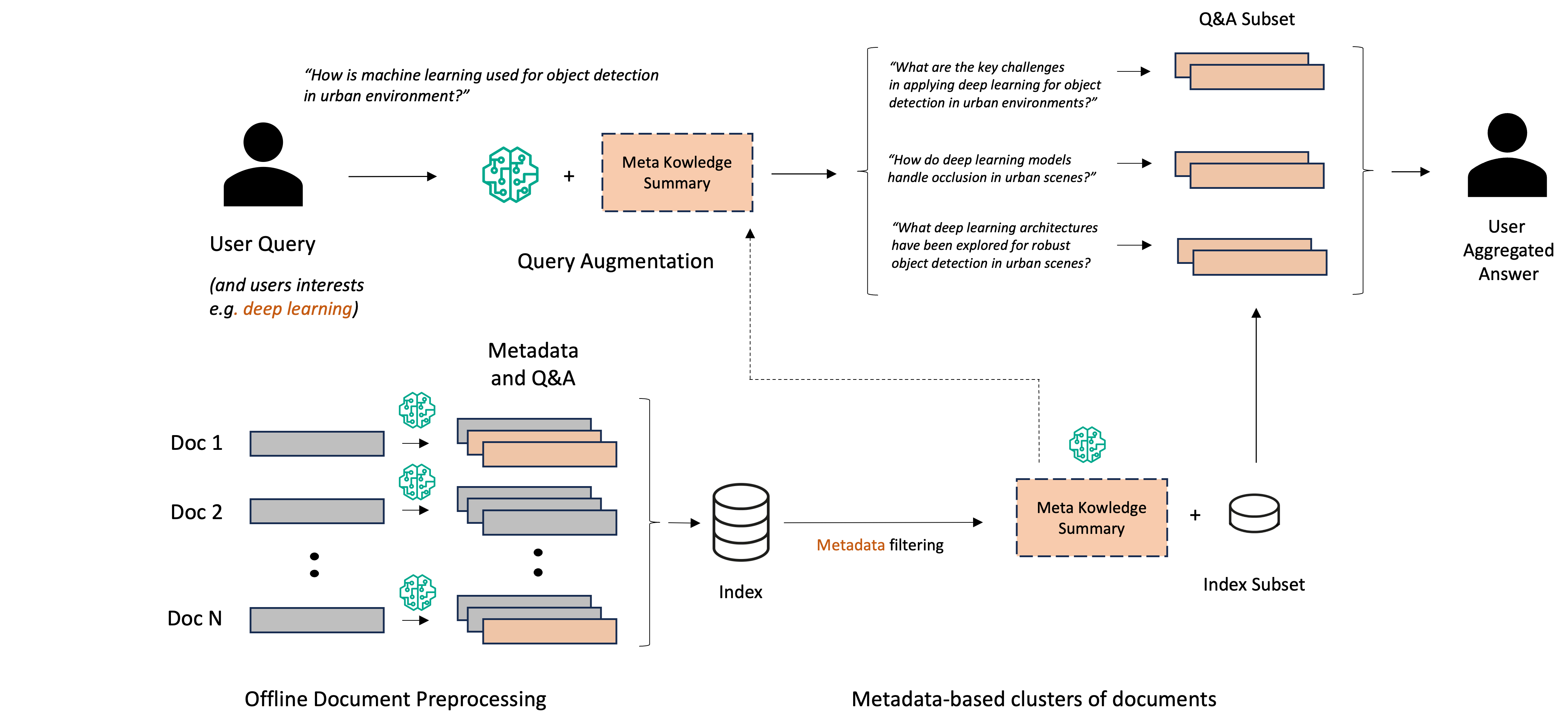}
  \caption{Data-centric workflow for Retrieval Augmented Generation (RAG) systems. Prior to inference, documents are augmented with Claude 3, and clustered into metadata-based sets of synthetic questions and answers for personalized downstream retrieval. Meta Knowledge Summaries are used to guide the query augmentation step with clusters information.}
  \Description{Test}
  \label{fig:teaser}
\end{teaserfigure}


\maketitle

\section{Introduction}
Retrieval Augmented Generation (RAG) is a standard technique used to augment Large Language Models (LLMs) with the capability to integrate contextually relevant, time-critical, or domain-specific information without altering the underlying model weights. This approach is particularly effective for knowledge-intensive tasks where proprietary or timely data is required to guide the language model’s response, and has become an attractive solution for reducing model hallucinations and ensuring alignment with the latest and most relevant information for the task at hand. In practice, RAG pipelines consist of several modules structured around the traditional retrieve-then-read framework \cite{lewis2020retrieval}. Given a user question, a retriever is tasked with dynamically searching for related document chunks and providing them as context for the LLM to predict the answer, rather than relying solely on pre-trained model knowledge (also known as in-context learning). A simple, yet powerful and cost-effective retriever framework involves using a dual-encoder dense retrieval model to encode both the query and the documents individually into a high-dimensional vector space, and computing their inner product as a measure of similarity \cite{karpukhin2020dense}. \

However, several challenges specifically hinder the quality of the knowledge augmented context. First, knowledge base documents may contain substantial noise, either intrinsic for the task at hand or as a result of the lack of standardization across the documents of interest (from various documents layouts or formats such as .pdf, .ppt, .wordx, etc.). Second, little to no human-labelled information or relevance labels are typically available to support the document chunking, embedding, and retrieval processes, making the overall retrieval problem a largely unsupervised approach and challenging to personalize for a given user. Third, chunking and separately encoding long documents pose a challenge in extracting relevant information for the retrieval models \cite{gao2023retrieval}. Indeed, document chunks do not conserve the semantic context of the entire document, and the larger the chunk, the less precise the context of the chunk is maintained for further retrieval. This makes the choice of document chunking strategy non-trivial for a given use-case, although critical for the quality of the subsequent steps due to potentially significant information loss. Fourth, user queries are typically short, ambiguous, may contain vocabulary mismatches, or are complex enough to require multiple documents to address, making it generally difficult to precisely capture the user’s intents and subsequently identify the most appropriate documents to retrieve \cite{zhu2023large}. Finally, there is no guarantee that the relevant information is localized in the knowledge base, but rather spread across multiple documents. As a result, domain expert-level usage of the knowledge base is made dramatically more challenging with automated information retrieval systems. Such high level reasoning across the knowledge base is a yet unsolved problem and constitutes the basis for recent LLM-based retrieval agent frameworks research. \

In this work, we are interested in cases where user queries require the information search to be specific to users interests or profile, are ambiguous, and require high level reasoning across documents (for example: \textit{“What challenges are associated with applying machine learning for marketing?"}), making recall, specificity, and depth our metrics of interest. To improve the search results performance across those metrics, query augmentation has been a widely used technique in both traditional Information Retrieval (IR) use cases such as e-commerce search \cite{peng2024large}, as well as in the more recent RAG frameworks leveraging LLMs \cite{gao2022precise}. Query augmentation consists in explicitly rewriting or extending the original user query into one or more tailored queries that better match search results, alleviating issues related to query underspecification. This adjustment adds a module to the RAG framework and transforms it into the more sophisticated rewrite-then-retrieve-then-read workflow. Leveraging their vast underlying parametric world knowledge, LLMs constitute a fitting choice to understand and enhance users queries, subsequently boosting the relevance of the retrieve step \cite{gao2022precise,ma2023query,mackie2023generative,mackie2023grm,shen2023large,jagerman2023query,srinivasan2022quill}. Our approach introduces a new data-centric RAG workflow, \textit{prepare-then-rewrite-then-retrieve-then-read} (PR3), where each document is processed by LLMs to create both custom metadata tailored to users characteristics and QA pairs to unlock new knowledge base reasoning capabilities through query augmentation. Our data preparation and retrieval pipeline mitigates the information loss inherent to large document chunking and embedding, as only QA are being encoded instead of documents chunks, while acting as a noise filtering approach for both noisy and irrelevant documents for the task at hand. By introducing metadata-based clusters of QAs and Meta Knowledge Summary, our framework conditionally augment the initial user query into multiple dedicated queries, therefore increasing the specificity, breadth, and depth of the knowledge base search (see Figure 1). The proposed out-of-the-shelve methodology is easily applicable to new datasets, does not rely on manual data labelling or model fine-tuning, and constitutes a step towards autonomous, agent-based documents database reasoning with LLMs, for which the literature remains limited to date \cite{zhu2023large}. 

\section{Related Work}
Our work integrates concepts from methodologies that generate QA from document collections for downstream fine-tuning of either LLMs or encoder models, with techniques leveraging query augmentation to boost the performance of retrievers in RAG pipelines. Below, we outline related work relevant to these two areas of RAG enhancement.

\subsection{RAG Enhancement with Fine-tuning}
Methodologies that aim at improving RAG pipelines based on fine-tuning generally constitute a higher barrier to entry for performing both the initial parameters update, and the maintenance of the accuracy of the model over time to new documents. They require careful data cleaning and (often manual) curation, and manual iterations across training hyperparameters sets to successfully adapt the model to the task at hand without causing catastrophic forgetting of the pre-trained model knowledge \cite{luo2023empirical}. In addition, model tuning may not be sustainable for frequent knowledge-base updates, and represents a generally higher cost due the underlying requirements of compute resources, despite the recent development of parameter-efficient fine-tuning (PEFT) techniques \cite{hu2021lora, dettmers2024qlora}. 

In e-commerce retrieval frameworks, TaoBao created a query rewriting framework based on company logs and rejection sampling to fine-tune a LLM in a supervised fashion, without QA generation. They further introduced a new contrastive learning methods to calibrate query generation probability to be aligned with desired search results, leading to a significant boost of merchandise volume, number of transactions and unique visitors \cite{peng2024large}. As an alternative, reinforcement learning methods based on black-box LLM evaluation has also been leveraged to train a smaller query-rewriter LLM, which showed a consistent performance improvement in open-domain and multiple-choice questions and answers (QA) in web search \cite{ma2023query}. Reinforcement learning-based approaches, however, are subject to more instability during the training phase, and require a careful investigations of the trade-offs between generalization and specialization among downstream tasks \cite{ma2023query}. Other approaches have focused on specifically improving the embedding space between the user query and the documents at hand, rather than augmenting the query itself. Authors of InPars \cite{bonifacio2022inpars} augmented their documents knowledge base by generating synthetic questions and answers pairs in a unsupervised fashion, and subsequently used them to fine-tune a T5 base embedding model. They showed that using the fine-tuned embedding model followed by a neural reranker such as ColBERT \cite{khattab2020colbert} outperformed strong baselines such as BM25 \cite{robertson2009probabilistic}. 

Most recently, other types of approaches have been developed to improve the end-to-end pipeline performance, such as RAFT \cite{zhang2024raft}, which consists in specifically training a reader to differentiate between relevant and irrelevant documents, or QUILL \cite{srinivasan2022quill} that aims at entirely replacing the RAG pipeline using RAG-augmented distillation training of another LLM.

\subsection{RAG Enhancement without Fine-tuning}

As an alternative to fine-tuning LLMs or encoder models, query augmentation methodologies have been developed to increase the performance of the retrievers by transforming the user query pre-encoding. These approaches can further be classified into two categories: either leveraging a retrieval pass through the documents, or zero-shot (without any example document). 

Among the zero-shot approaches, HyDE \cite{gao2022precise} introduced a data augmentation methodology that consists in generating an hypothetical response document to the user query by leveraging LLMs. The underlying idea is to bring closer the user query and the documents of interest in the embedding space, therefore increasing the performance of the retrieval process. Their experiments showed performance comparable to fine-tuned retrievers across various tasks. The generated document, however, is a naïve data augmentation in the sense that it does not change given the underlying embedded data for the task at hand, such that it can lead to performance decrease in multiple situations, for there is inevitably a gap between the generated content and the knowledge base. Alternatively, methodologies have been proposed to perform an initial pass through the embedding space of the documents first, and subsequently augment the initial query to perform a more informed search. These Pseudo Relevance Feedback (PRF) \cite{mackie2023generative} and Generative Relevance Feedback (GRF) modeling approaches \cite{mackie2023grm} are typically dependent on the quality of the most highly-ranked documents used to first condition their query augmentation to, and are therefore prone to significant performance variation across queries, or may even forget the essence of the original query. 

\section{Methodology}

In both RAG pipeline enhancements approaches cited above, the retrievers are generally unaware of the distribution of the target collection of documents despite an initial pass through the retrieval pipeline.
In our proposed framework, for each document and prior to inference, we create a set of dedicated metadata, and subsequently generate guided QA spanning across the documents using Chain of Thoughts (CoT) prompting with Claude 3 Haiku \cite{anthropic2024claude,wei2022chain}. The synthetic questions are then encoded, and the metadata used for filtering purposes. For any user-relevant combination of metadata, we create a Meta Knowledge Summary (MK Summary), leveraging Claude 3 Sonnet, which consists in a summarization of the key concepts available in the database for a given filter. At inference time, the user query is dynamically augmented by relying on the personalized MK Summary given the metadata of interest, therefore providing tailored response for this user. By doing so, we provide the retriever with the capability to reason across multiple documents which may have required multiple retrieval and reasoning rounds otherwise. Our goal is to ultimately increase the quality of the end-to-end retrieval pipeline across multiple metrics, such as depth, coverage, and relevancy, by enabling complex reasoning across the database through tailored searches and the leverage of meta knowledge information. Importantly, our approach does not rely on any model weights update, and may very well be combined with any fine-tuning of either the language or the encoding models to any domain to further improve the performance of the end-to-end RAG pipeline \cite{gupta2024rag}. We represent our methodology pipeline in Figure 1, and describe the synthetic QA generation process and the concept of MK Summary below. 

\subsection{Datasets}
Our public benchmark use case comprises a dataset of 2,000 research papers from 2024, collated using the arXiv API. This dataset represents a diverse spectrum of research in statistics, machine learning, artificial intelligence, and econometrics\footnote{The dataset was filtered using the following categories on the Arxiv API: "stat.ML", "stat.TH", "stat.AP", "stat.ME", "math.ST", "cs.AI", "cs.LG", "econ.EM". Thank you to arXiv for use of its open access interoperability.}, for a total of approximately 35M tokens.

\subsection{Synthetic QA Generation}

First, for each document, we generate a set of metadata and subsequent QA using CoT prompting (see Appendix A). The prompt aims at creating a list of metadata by classifying the documents into a predefined set of categories (such as research field, or applications types for our research papers benchmark). Relying on these metadata, we generate a set of synthetic questions and answers, using teacher-student prompting and assess the knowledge of the student on the document. We specifically leverage Claude 3 Haiku for its long-context reasoning abilities and potential to create synthetic QA pairs with context spanning across documents. The generated metadata serve both as filtering parameters for the augmented search, and to select the synthetic QA used for the users queries augmentation in the form of meta knowledge information (MK Summary). In addition, the synthetic QA are used for the retrieval, but only questions are vectorized for downstream retrieval. For our public scientific research papers use case, a total of 8657 QA were generated from the 2000 research documents, accounting for 5 to 6 questions for 70\% of cases, and 2 questions in 21\% of cases. Examples of synthetic questions and answers are provided in Appendix B. The total number of token generated as parts of the processing step amount to approximately 8M output tokens, corresponding to a total of \$20.17 for the entire processing pipeline of all 2000 documents (including input tokens) using Amazon Bedrock \cite{pricingbedrock}. We investigated the redundancy of the generated QA across the document using hierarchical clustering on the embedding space of the questions using e5-mistral-7b-instruct \cite{wang2023improving}, but did not de-duplicated the generated QA for our use cases due to the low QAs overlap. QA Filtering can be application and metadata specific, and other high-dimensional approaches such as Determinantal Point Processes (DPP) \cite{kulesza2012determinantal} are left for future work, together with the automated discovery of metadata topics and self-correcting QA-generation \cite{pan2023automatically}.

\subsection{Generation of Meta Knowledge Summary}

For a given combination of metadata, we create a Meta Knowledge Summary (MK Summary) aiming at supporting the data augmentation phase for a given user query. For our research papers use case, we limited our metadata to the specific field of research (such as reinforcement learning, supervised vs unsupervised learning, bayesian methods, econometrics, etc.) identified during the document processing phase by Claude 3 Haiku. For this research, we create the MK Summary by summarizing the concepts across a set of questions tagged with the metadata of interest using Claude 3 Sonnet. An alternative left for future work is that of prompt tuning to optimize for the content of the summary prompt \cite{tam2022parameter}.

\subsection{Augmented Generation of Queries and Retrieval}
Given a user query and a set of pre-selected metadata of interests, we retrieve the corresponding pre-computed MK Summary and use it to condition the user query augmentation into the database subset. For our research paper benchmark, we created a set of 20 MK Summary corresponding to research fields (e.g. deep learning for computer vision, statistical methods, bayesian analysis, etc.), relying on the metadata created in the processing phase. We leverage the "plan-and-execute" prompting methodology to address complex queries, reason across documents, and ultimately improve the recall, precision, and diversity of the provided answers \cite{sun2023pearl}. For example, for a user query related to the Reinforcement Learning research topic, the pipeline will first retrieve the meta knowledge (MK Summary) about Reinforcement Learning of the database, augment the user query into multiple sub queries based on the content of the MK Summary, and perform a parallel search in the filtered database relevant for manufacturing questions. For this purpose, the synthetic Questions are embedded, and replace the original documents chunk-based similarity matching, therefore mitigating the information loss due to document chunking discontinuity. Once the best match of a synthetic question is found, the corresponding QA are retrieved, together with the original document title. Only the document title, the synthetic question, and the answer are returned as a result of the retrieval. We use JSON formatting for downstream summarization performance. The final response of the RAG pipeline is obtained by providing the original query, the augmented queries, the retrieved context and few shot examples (see Figure 1).

\section{Evaluation}
\subsection{Generating Evaluation Queries}
To evaluate our data-centric augmented retrieval pipeline, we generated 200 questions for the arXiv dataset using Claude 3 Sonnet (see Appendix C). In addition, we compared our methodology against traditional document chunking, query augmentation with document chunking, and naïve (not using MK Summary) augmentation with the QA processing of the documents. As a comparison, we created documents chunks consisting of 256 tokens with 10\% overlap. For our use case, traditional chunking generated 69,334 documents chunks.

\subsection{Evaluation Metrics and Prompts}

Without relevance labels, we use Claude 3 Sonnet as a trusted evaluator \cite{anthropic2024claude} to compare the performance of all four benchmark methodologies considered: traditional chunking without any query augmentation, traditional document chunking with naive query augmentation, augmented search using our PR3 pipeline without MK Summary, and augmented search using our PR3 pipeline with MK Summary. The query augmentation prompt is provided in Appendix D. We use the custom performance metrics defined below directly in the prompt to compare the results of both the retrieval model and the final response on a scale from 0 to 100. An example of Claude 3 Sonnet comparison answer is provided in Appendix E.\\

- \textbf{Recall}: evaluates the coverage of key, highly relevant information contained in the retrieved documents

- \textbf{Precision}: evaluates the ratio of relevant documents against irrelevant ones

- \textbf{Specificity}: evaluates how precisely focused the final answer is on the query at hand, with clear and direct information that addresses the question

- \textbf{Breadth}: evaluates the coverage of all relevant aspects or areas related to the question, providing a complete overview

- \textbf{Depth}: evaluates the extent to which the final answer provides a thorough understanding through detailed analysis and insights into the subject

- \textbf{Relevancy}: evaluates how well-tailored the final answer is to the needs and interests of the audience or context, focusing on providing directly applicable and essential information while omitting extraneous details that do not contribute to addressing the specific question

\section{Results}
We considered 4 cases for the evaluation of our retrieval pipeline: (1) traditional document chunking, without any augmentation, (2) traditional document chunking and augmentation, (3) QA-based search and retrieval, with naïve augmentation (our first proposal), and (4) QA-based search and retrieval, with the use of MK summary (our second proposal). For a single query, the computational latency of the end-to-end-pipeline amounts to ~20-25 seconds.

\subsection{Retrieval and End-to-end Evaluation Metrics}
For each of the synthetic user queries generated, we ran a comparison prompt that includes the context retrieved as part of each approach, together with their final answers. We prompted Claude 3 Sonnet to rate each of the metrics on a scale from 0 to 100, together with a justification text. An example of the evaluation response is provided in Appendix E. The obtained metrics are then averaged across all queries and displayed below in Figure 2. We observe a clear benefit across all metrics but the precision of the retrieved documents by our two proposed QA-based methodologies. The lack of strong improvement over the precision metric is consistent with the usage of a single encoding model and show that few documents are considered completely irrelevant. Specifically, we note a significant performance boost in both the breadth and the depth of the final LLM response. This result shows that the MK Summary is providing additional information that is leveraged by the query augmentation step. Finally, the contribution of the MK summary to the conditioning of the search itself appears statistically significant across all metrics but the precision of the retriever ($p < 0.01$ between the augmented QA search and the MK-Augmented QA search)(see Table 1). We observe that the the proposed methodology significantly improves the breadth of the search (by more than 20\%, compared to traditional naïve search with chunking approaches), which aligns to the intuition that our proposal allows for effectively synthetizing more information from the content of the database, and leveraging its content more extensively.

\begin{table}[h]
\caption{Performance benchmark across 200 Synthetic user queries}
\centering
\label{tab:combined}
\small
\resizebox{0.5\textwidth}{!}{%
\begin{tabular}{cccc} 
    \toprule
    \textbf{Public Research Benchmark} & Recall (\%) & Precision (\%) & Specificity (\%)  \\
    \midrule
    Naïve Search with Chunking & 77.76 & 86.91 & 71.51 \\
    Augmented Search with Chunking & 82.27 & 87.09 & 74.86 \\ 
    Augmented QA Search & 86.33 & 90.04 & 79.64 \\ 
    MK-Augmented QA Search & 88.39 & 90.40 & 83.03 \\ 
    \bottomrule
\end{tabular}
}
\resizebox{0.5\textwidth}{!}{%
\begin{tabular}{cccc} 
    \toprule
    \textbf{Public Research Benchmark} & Breadth (\%) & Depth (\%)& Relevancy (\%)  \\
    \midrule
    Naïve Search with Chunking & 67.32 & 65.62 & 81.51 \\
    Augmented Search with Chunking & 79.77 & 72.41 & 85.08 \\ 
    Augmented QA Search & 84.55 & 78.08 & 88.92 \\ 
    MK-Augmented QA Search & 87.09 & 80.84 & 90.22 \\ 
    \bottomrule
\end{tabular}
}
\end{table}

\begin{figure}[tbp]
  \centering
  \includegraphics[width=0.4\textwidth]{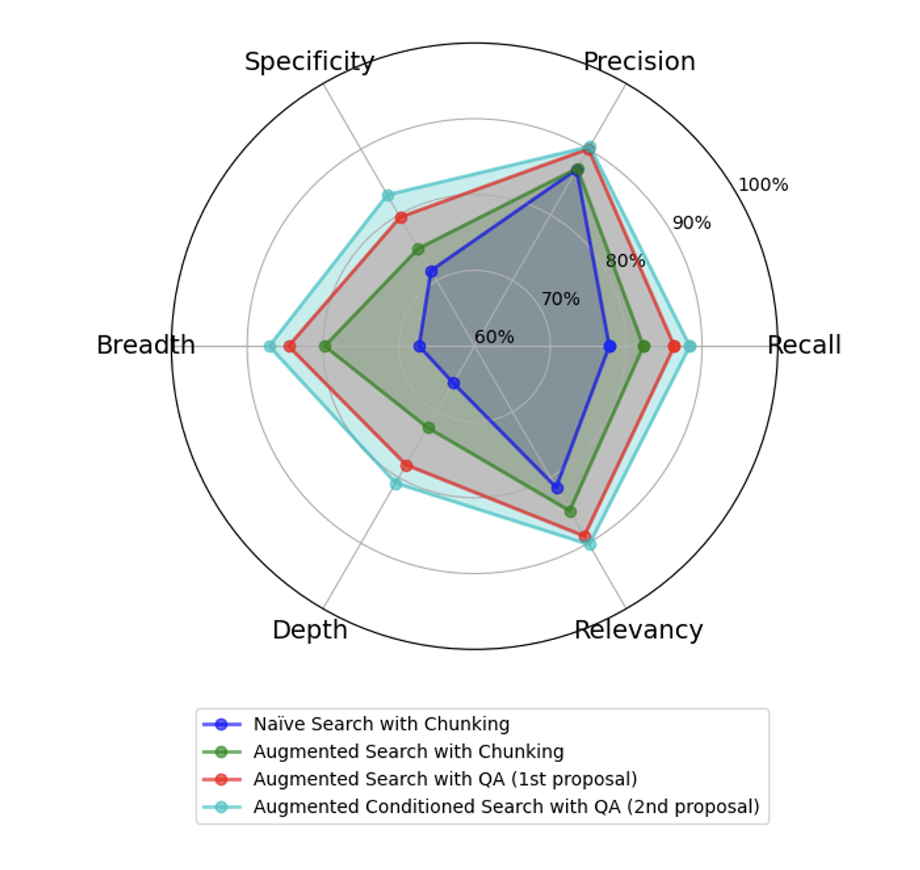}
  \caption{Summary of results obtained without any augmentation and with naïve document chunking, compared to results obtained using synthetic QA (proposal 1) and MK Summary (proposal 2).}
  \label{fig:fig1}
\end{figure}

\section{Conclusion and Discussion}
We proposed a new data-centric RAG workflow which leverages synthetic QA generation instead of the traditional document chunking framework, and a query augmentation-based approach based on high level summary of the content metadata-based clusters of documents to improve the accuracy and quality of the end-to-end LLM augmentation pipeline. Our methodology significantly outperforms traditional RAG pipelines relying on document chunking and naïve user query augmentation. We introduced the concept of MK Summary to further boost the zero-shot search augmentation in the knowledge base, which subsequently increased the performance of the end-to-end RAG pipeline in our test case. In essence, our methodology improves on simple semantic matching information retrieval in the encoding vector space of the documents, where we allow for more diverse but highly relevant documents search, therefore providing more well-rounded, domain expert-level, and comprehensive answers to the user query. On all metrics considered, recall, precision, specificity, breadth, depth, and relevancy, the proposed approach improved on state-of-the-art work. Finally, the approach is cost-effective, costing \$20 for 2000 research papers. As a limitation, while we recognize the difficulty in crafting a set of metadata prior to document processing, the metadata generation can become an iterative approach to generate the metadata upon discovery. In addition, we left multi-hop iterative searches and improvement of the summary of the clustered knowledge base for future work.

\bibliographystyle{unsrt}
\bibliography{ref}

\small

\appendix

\section{Appendix}
\subsection{Prompts}
\begin{lstlisting}
You are an helpful research assistant, preprocessing {document_types} for {users_types} to use later on. 
You are provided with a document and a list of questions that aims at extracting key knowledge from this document. Please stricly follow the format below to answer (no introduction or finishing sentences).

First, answer the following questions with a single Yes or No only:
1. The paper can be clearly categorized into one or multiple research field(s) (exclusively from: {text_categories}), yes or no?:
2. The paper is mostly an applied research paper (versus mostly theoric), yes or no?:
3. The paper is referencing a Github repository, yes or no?
4. The paper contains mathematical reasoning, yes or no?:
5. The paper mentions a specific application to an industry company, yes or no?:
6. The paper uses evaluations metrics to benchmark their methods, yes or no?:
Answer the following questions with a python list only, or return an empty python list:
1. If the paper can be clearly categorized into one or multiple research fields, list the fields (3 max):
2. If the paper is mostly an applied research paper, list the application fields (3 max):
3. If the paper references one or more Github repository, list their urls (2 max):
4. If the paper contains mathematical reasoning, list the name(s) of the theorem(s) being used (3 max):
5. If the paper mentioned a specific application to an industry company, list the companies (3 max):
6. If the paper use evaluations metrics to benchmark their methods, list the names of the metrics (5 max):

Your answer must look like the following (no introduction sentence):
1. Yes
2. No
etc.

1. ['a','b']
2. []
etc.

Then, please act as an expert scientists and formulate both general (general understanding) and precise questions (incl. specific findings or limitations) from the content of the document to assess the knowledge of other highly knowledgeable scientists about the topic of this document. 

Scientists that will answer the questions do not know the document. Please do not explicitly refer to "the text" or the name of the document in the questions. 
Each questions and answers pairs must be self-contained (make sure to give enough context) and independent from other pairs. 

Please formulate as many questions as possible covering as much content as possible, and avoid bullet points within answers.

Stricly follow the format of the final questions and answers below, presenting all responses, lists, questions, then all answers:

Questions:
1. ...
2. ...
etc. 

Answers: 
1. ...
2. ...
etc. 

Please find below the text, for which the title is {doc_title}:

[Text]
{doc_content}
[/Text]
"""
\end{lstlisting}

\section{Example of Synthetic QA}

We present below samples of synthetic QA generated from the 2,000 research papers. The questions and answers were generated using the metadata and QA generation prompt (Appendix A).\\

- Question \#1: "What are the researchers' conjectures about why the future narrative prompting approach seemed to unlock improved predictive capabilities in ChatGPT-4 compared to direct prompting?" (from \cite{pham2024chatgpt}) 

- Answer \#1: "The researchers conjecture that the narrative prompting approach may leverage ChatGPT-4's capacity for "hallucinatory" narrative construction, allowing it to more effectively synthesize and extrapolate from its training data to make more accurate predictions, compared to the direct prompting approach. This suggests the models' creative abilities can be strategically utilized for predictive tasks."\\

- Question \#2: "How does the distribution of preference datasets influence the rate of model updates and the training accuracy when using the Direct Preference Optimization (DPO) objective? What are the implications for practical alignment approaches?" (from \cite{im2024understanding})

- Answer \#2: "The paper provides theoretical analysis showing that the distribution of preference datasets, characterized by "preference distinguishability", has a strong influence on the rate of model updates and training accuracy under the DPO objective. Behaviors with higher distinguishability exhibit faster loss reduction and parameter updates. This can lead to an intricate prioritization effect, where DPO tends to prioritize learning behaviors with higher distinguishability, potentially deprioritizing less distinguishable but crucial behaviors."

\section{Benchmark User Queries}

We present below a sample of 200 benchmark research queries generated with Claude 3 Sonnet. \\

- Field of Research (metadata) \#1 : Reinforcement Learning

- Synthetic Benchmark query \#1: How is reinforcement learning used in adaptive traffic management systems?\\

- Field of Research (metadata) \#2 : Econometrics

- Synthetic Benchmark query \#2: How is econometrics applied in understanding consumer demand elasticity?\\

- Field of Research (metadata) \#3 : Statistical Anomaly Detection

- Synthetic Benchmark query \#3: How are anomalies identified in unstructured data like images and videos?\\

- Field of Research (metadata) \#4 : Supervised vs Unsupervised Learning

- Synthetic Benchmark query \#4: How are supervised learning techniques adapted for complex, high-dimensional data?\\

- Field of Research (metadata) \#5 : AI in Autonomous Vehicles

- Synthetic Benchmark query \#5: What advancements have been made in integrating sensor data using AI in autonomous vehicles?\\

\section{Query Augmentation Prompt}

\begin{lstlisting}
You are an assistant for scientists. You will be provided with a user question. 
Your goal is to generate questions for scientists to asks the literature and prepare themselves to answer it.
To generate the questions, you can rely on the summary of the database at hand, provided below as [DatabaseSummary]
Please generate as much relevant questions as possible (maximum 5) for a strategic answer.
Remember, scientists will use these questions to search the literature.

It is better to generate more simple questions than fewer complex questions.

[DatabaseSummary]
{mk_summary}
[/DatabaseSummary]

[HumanQuery]
{user_query}
[/HumanQuery]

Please only reply with numbered questions, stricly follow this format. There is no need for introduction context or conclusion text before or after questions.

Example:

1. ...
2. ...
3. ...
...
N. ...
"""
\end{lstlisting}
\section{LLM Comparison Response}

\begin{lstlisting}
**Scientist4 Response**

**Recall:** 90/100 - The response covers a comprehensive range of information related to data privacy concerns in autonomous vehicles, including data collection, privacy risks, system design considerations, legal frameworks, and ethical considerations.

**Precision:** 95/100 - The information provided is highly relevant and focused on addressing the question, with minimal irrelevant details.

**Specificity:** 85/100 - The response provides specific examples and details, such as mentioning the GDPR, data anonymization techniques, and ethical principles like transparency and accountability.

**Breadth:** 90/100 - The response covers a wide range of aspects related to data privacy concerns, including technical, legal, and ethical considerations, providing a comprehensive overview.

**Depth:** 85/100 - The response offers in-depth insights and analysis into the complexities involved in addressing data privacy concerns in autonomous vehicles, demonstrating a thorough understanding of the subject matter.

**Relevancy:** 95/100 - The information provided is highly relevant and directly addresses the key aspects of data privacy concerns in autonomous vehicles, while also providing additional context and considerations.
\end{lstlisting}

\end{document}